\documentclass[aps,prl,reprint,showpacs,groupedaddress,amsmath,amssymb,floatfix,tightenlines]{revtex4-1}
\usepackage{graphicx}
\usepackage{graphics}

\begin{document}



\title{Order-disorder transition and alignment dynamics of a block copolymer under high magnetic fields by in situ x-ray scattering}


\author{Manesh Gopinadhan}
\email[]{manesh.gopinadhan@yale.edu}
\affiliation{Department of Chemical and Environmental Engineering, Yale University,
New Haven CT 06511}
\author{Pawe{\l} W. Majewski}
\email[]{pawel.majewski@yale.edu}
\affiliation{Department of Chemical and Environmental Engineering, Yale University,
New Haven CT 06511}
\author{Youngwoo Choo}
\email[]{youngwoo.choo@yale.edu}
\affiliation{Department of Chemical and Environmental Engineering, Yale University,
New Haven CT 06511}
\author{Chinedum O. Osuji}
\email[]{chinedum.osuji@yale.edu}
\affiliation{Department of Chemical and Environmental Engineering, Yale University,
New Haven CT 06511}


\date{\today}

\begin{abstract}
We examine the influence of magnetic fields on the order-disorder transition (ODT) in a liquid crystalline block copolymer. This is motivated by a desire to understand the dynamics of microstructure alignment during field annealing as potentially dictated by selective destabilization of non-aligned material. Temperature resolved scattering across the ODT and time-resolved measurements during isothermal field annealing at sub-ODT temperatures were performed \textit{in situ}. Strongly textured mesophases resulted in each case but no measurable field-induced shift in $T_{ODT}$ was observed. This suggests that selective melting does not play a discernable role in the system's field response. Our data indicate instead that alignment occurs by slow grain rotation within the mesophase. We identify an optimum sub-cooling that maximizes alignment during isothermal field annealing. This is corroborated by a simple model incorporating the competing effects of an exponentially decreasing mobility and divergent, increasing magnetic anisotropy on cooling below $T_{ODT}$. The absence of measurable field-effects on $T_{ODT}$ is consistent with estimates based on the relative magnitudes of the field interaction energy and the enthalpy associated with the ODT.
\end{abstract}


\maketitle


The use of external fields to direct self-assembly in soft matter is predicated on anisotropic field interactions that produce a well-defined free energy minimum in the aligned state \cite{darling2007dsa}. Due to the absence of dielectric breakdown concerns and their space pervasive nature, magnetic fields  are particularly well suited for this purpose. They couple to the magnetic susceptibility anisotropy, $\Delta\chi$, and effectively permit arbitrary control of texture in appropriately field-responsive materials, even in complex geometries \cite{majewski2012magnetic}. Typical block copolymers (BCPs) such as poly(styrene-\textit{b}-methyl methacrylate) exhibit vanishingly small anisotropies with $\Delta\chi$ $\mathcal{O}(10^{-9})$ (SI units). This is due to the random coil nature of the chains and the similarity in $\chi$ of the component species. The presence of rigid anisotropic moieties such as mesogens in liquid crystalline (LC) mesophases \cite{Maan2003mesogene,Osuji_Macromol2004,Hamley2004,Segalman_magnetic_2007,Xu_magnetic_align_FaradayDisc2009,gopinadhan2012magnetic} as well as extended chains in semi-crystalline polymers \cite{kimura2003study,Thurn-Albrecht_magnetic_alignment2005} and surfactants \cite{Chmelka_JACS1997,majewski_SoftMatter2009,wilmsmeyer2011switchable} gives rise to $\Delta\chi$ $\mathcal{O}(10^{-6})$ which is sufficient to drive alignment at field strengths of a few tesla. Magnetic nanoparticle inclusions can also be leveraged to align soft matter systems \cite{vallooran2011macroscopic}.

The mechanism by which the transition occurs from disordered melts to ordered field-aligned mesophases in LC BCPs is a subject of much interest in soft matter physics. This stems in part from the considerable potential for aligned BCPs to address a broad range of applications \cite{hamley2003nsm}. At issue is the question of whether alignment is dominated by grain rotation or by selective field-induced destabilization of non-aligned material. This is linked to the broader topic in materials physics of field effects on first order phase transitions especially as such effects dictate the potential role of selective melting. Despite the considerable potential of field alignment to enable a wide range of BCP-based technologies \cite{hamley2003nsm}, studies of magnetic field driven phase behavior and alignment dynamics in BCPs have been notably limited to date. This is in part due to the need for high field magnets and the experimental difficulty associated with measurements in the presence of large fields.

Here we report on \textit{in situ} magnetic field studies of an LC BCP utilizing a specialized small angle x-ray scattering (SAXS) lab instrument coupled to a high field magnet. Temperature resolved measurements show no discernible field effects on the order-disorder transition (ODT) of the system. Time resolved experiments on pre-aligned materials show that alignment at sub-ODT temperatures occurs via grain rotation with slow kinetics. The response of the system is critically limited by the mesophase viscosity such that alignment can only be advanced by residence in a small temperature window near $T_{ODT}$. This residence produces a weakly aligned system which thereafter transitions to a strongly aligned state on cooling, even in the absence of the field. To the best of our knowledge, this work represents the first such \textit{in situ} systematic examination of BCP phase behavior and alignment dynamics under magnetic fields. We provide simple models which correctly capture the absence of measurable field-effects on the ODT and the existence of an optimum sub-cooling for the maximization of orientation order parameters in isothermal experiments.

\begin{figure}[t]
\centering
\includegraphics[width=85mm, scale=1]{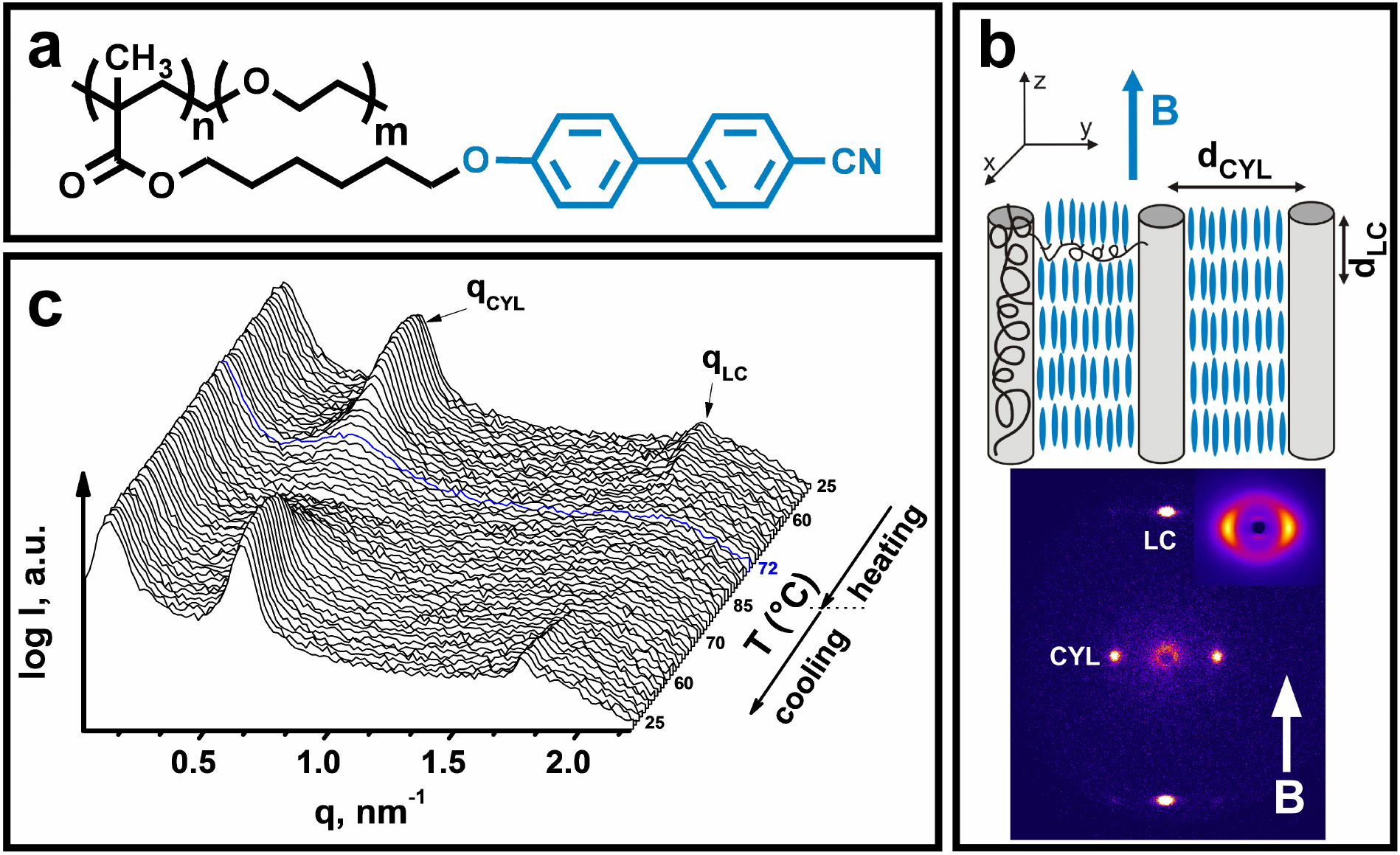}
\caption{a: Chemical structure of the LC BCP. b: Schematic showing planar anchoring of mesogens at the PEO/PMA interface. Mesogens and PEO cylinders align parallel to the field as shown by 2D SAXS and WAXS (inset). c: Temperature resolved SAXS at zero-field. The blue trace marks $T_{ODT}$}
\label{odt_scheme}
\end{figure}

The polymer studied is poly(ethylene oxide-\textit{b}-methacrylate), PEO-PMA/LC, in which the MA block bears side-attached cyanobiphenyl mesogens (Polymersource), Fig. \ref{odt_scheme}a. The molecular weight $M_n$ is 10.4 kg/mol with PEO weight fraction $f_{PEO}$=0.23 and polydispersity index $\approx$ 1.1. The system forms hexagonally packed cylinders of PEO in the PMA/LC matrix, with a d-spacing, \mbox{$d_{cyl}$}=9.6 nm. Samples were prepared by solvent casting from 5 wt.\% dimethyl formamide solutions to provide 3 mm diameter discs of 1-2 mm thickness. SAXS was conducted on a customized Rigaku instrument integrated with a 6 T cryogen-free magnet (AMI).

\begin{figure}
\centering
\includegraphics[width=85mm, scale=1]{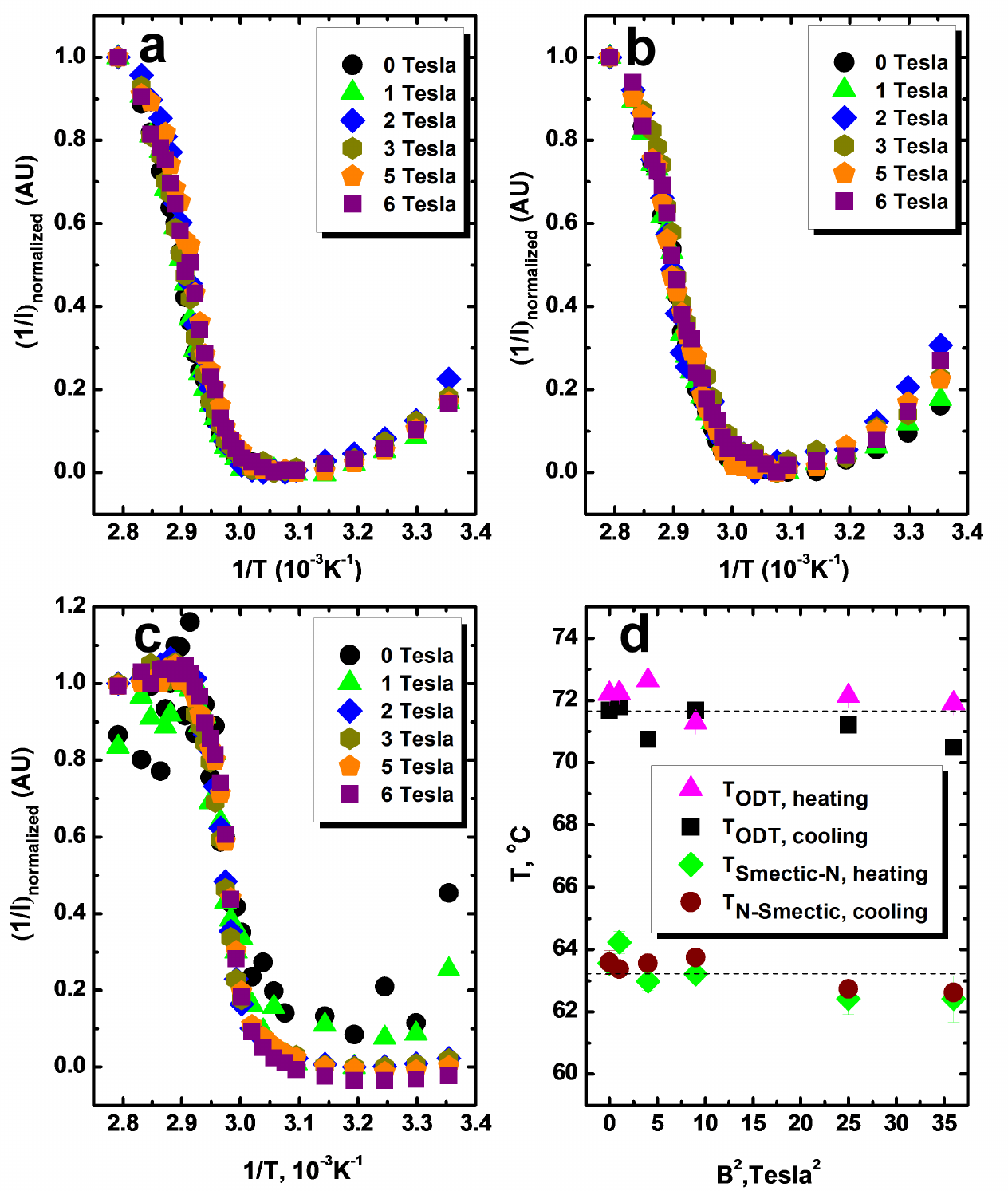}
\caption{Inverse peak intensity as a function of $1/T$ for BCP on heating (a) and cooling (b), and for smectic LC scattering on cooling (c). d: $T_{ODT}$ and $T_{SmA-N}$ as a function of $B^2$.}
\label{saxs_field}
\end{figure}

The phase behavior is typical of weakly segregated LC BCPs. Differential scanning calorimetry (DSC) (not shown) and zero-field SAXS, Fig. \ref{odt_scheme}c, show that the ODT at $\approx$ 72 $^{\circ}$C is effectively coincident with $T_{NI}\approx 69 ^{\circ}$C. Self-assembly here is driven by a change in block interaction on formation of the nematic phase, as also seen in other weakly segregated LC BCPs \cite{zheng1998phase}. A nematic to smectic A  transition occurs near 63 $^{\circ}$C. The mesogens exhibit planar anchoring at the interface between the PEO and PMA chains. Consequently, the cylindrical microdomains are parallel to the mesogens and the periodicity of the smectic layers, $d_{LC}$=3.5 nm, is orthogonal to that of the cylindrical microdomains, Fig. \ref{odt_scheme}b.  Plots of $I^{-1}(1/T)$ at q=1.8 and 0.65 nm$\mathrm{^{-1}}$ for the Smectic A layers and cylindrical microdomains are shown in Fig. \ref{saxs_field}. Samples were initially heated to $\approx$ 85 $^{\circ}$C, well above $T_{ODT}$ and then cooled at 0.1 $^{\circ}$C/min to 25 $^{\circ}$C under different field strengths, stopping every 1 $^{\circ}$C for 600 s for data collection. Data were fitted using a standard logistic function to determine $T_{ODT}$ and $T_{N-SmA}$, Fig. \ref{saxs_field}d. The results show there is no discernible influence of the field on $T_{NI}$ and $T_{SmA}$ beyond an uncertainty of roughly $\pm$0.5 $^{\circ}$C associated with the experimental apparatus. In all cases except the zero-field measurement, the cylinders were aligned parallel to the field direction. There was pronounced narrowing of the azimuthal intensity distributions for field strengths above 2 T, with a saturation of the peak width (FWHM) of roughly 7-8$^{\circ}$ and 4-5$^{\circ}$ for the microdomain and smectic layer reflections respectively \cite{Majewski_anisotropic_conductivity2010}.

The role of the field on $T_{NI}$ and thus the ODT can be considered through construction of a Clausius-Clapeyron relation incorporating the field interaction energy at constant pressure \cite{rosenblatt1981magnetic}. The normalized change in transition temperature is given by Eq. \ref{eq:field_ODT} where the susceptibility of the nematic state can be closely related to that of the isotropic state and the molecular anisotropy as $\chi_N=\chi_I+(2/3)\Delta\chi_m$, the field interaction energy difference is $(\chi_N-\chi_I) B^2/2\mu_0$ and $\Delta H_{NI}$ is the enthalpy associated with the transition.

\begin{equation}
\Delta T/T_{NI}=\frac{\Delta\chi_m B^2}{3\mu_0\Delta H_{NI}}
\label{eq:field_ODT}
\end{equation}

Using representative values of $\Delta H_{NI}$=1 J/g (from DSC), $\Delta\chi_m$=10$^{-6}$ and $\rho$=1 g/cm$^3$, a field strength of 6 T would impose a very small shift of $\Delta T\approx$ 4 mK, which is far below the resolution of the measurements. Such mK-scale shifts are consistent with prior work on electric \cite{helfrich1970effect} and magnetic field studies \cite{rosenblatt1981magnetic} on small molecule LCs. Intriguingly, Segalman \textit{et al.} observed a large 30 K shift in the ODT of a weakly segregated rod-coil BCP under application of a 7 T field \cite{Segalman_ODT2011}. It is likely that the small entropy changes associated with ODTs in rod-like systems results in this exceptionally large field effect by comparison with the present side-chain LC BCP and small molecule systems based on analogous cyanobiphenyl mesogens. Conversely Samulski \textit{et al.} examined a  system based on a bent-core mesogen with non-trivial transition enthalpies but identified a large 4 K shift in $T_{NI}$ and $T_{SmC-N}$ at 1 T. This was attributed to the presence of partially ordered clusters which couple to the field in this novel class of cybotactic materials \cite{francescangeli2011extraordinary}.

Pronounced field-induced changes in transition temperatures can drive selective melting whereby domains which are not in the energy minimizing orientation are destabilized, and thus shrink, or ``melt'', and are replaced by aligned material growing from the disordered melt \cite{koppi1992lamellae,amundson1994alignment,boker2002microscopic,Liedel2012}. This acts as an effective mechanism for mesophase alignment. In the present case, given the imperceptible field effects, we can conclude that while selective melting may occur, it does so only over a mK-scale window which cannot be easily leveraged as a viable means of structure control. Correspondingly, this mechanism has little consequence in the current experiments.

\begin{figure}
\centering
\includegraphics[width=85mm, scale=1]{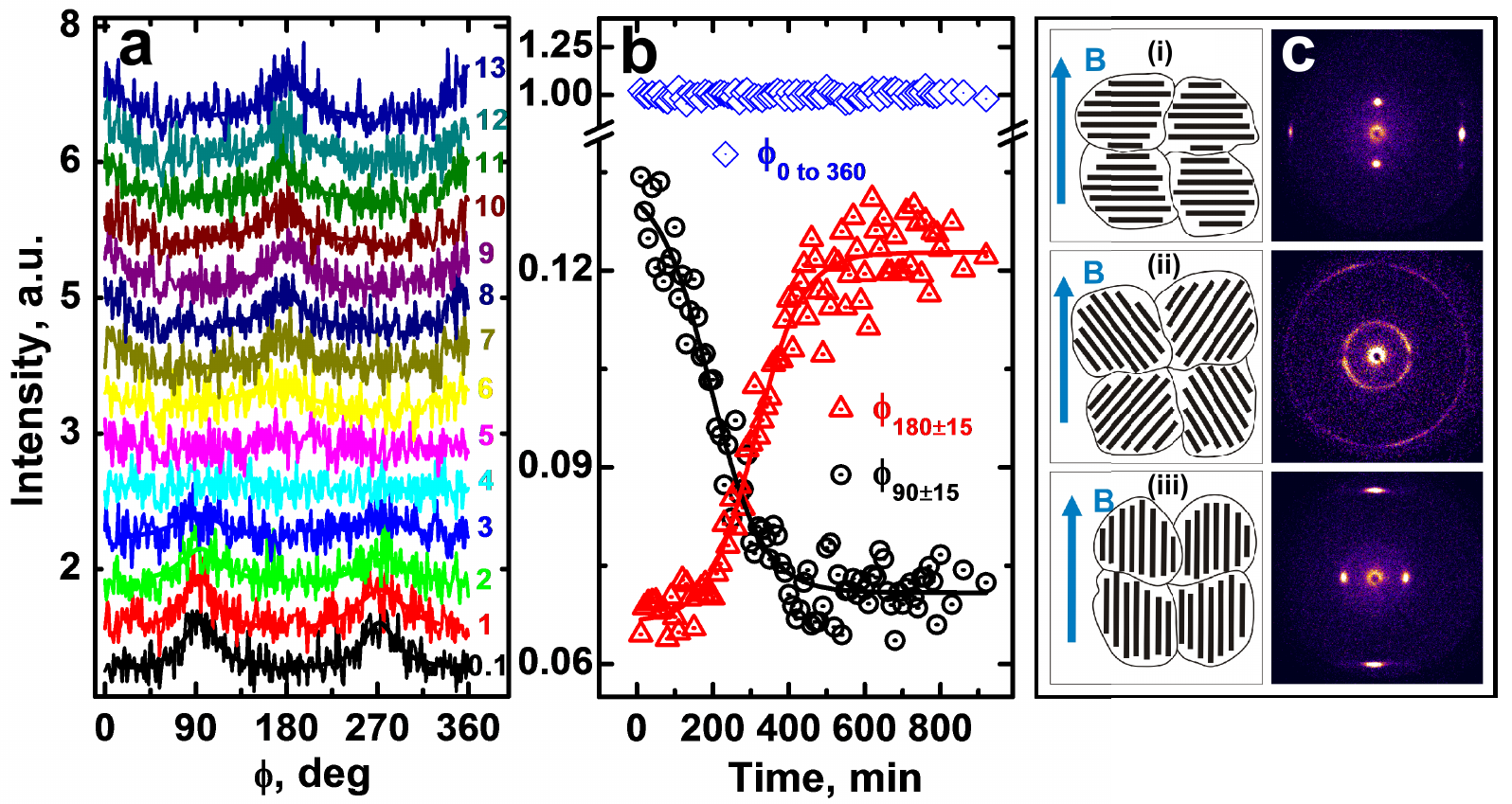}
\caption{a: Azimuthal intensity at selected times (hrs). b: Integrated peak intensities centered at $\phi$=90$^{\circ}$ and 180$^{\circ}$ with exponential fits (lines) and total intensity from 0 to $2\pi$ at q=0.7 nm$^{-1}$. c: 2D SAXS and grain reorientation schematics at $t=0$ (25 $^{\circ}$C), $t=250$ (50 $^{\circ}$C) and $t=900$ min (25 $^{\circ}$C).}
\label{kinetics}
\end{figure}

The alignment kinetics were probed using isothermal field annealing to monitor microstructure re-orientation at sub-ODT temperatures. A sample was initially well aligned by slowly cooling across the ODT to 25 $^{\circ}$C at 0.1 $^{\circ}$C/min at 6 T. The field was ramped to 0 T and the sample was physically rotated by 90$^{\circ}$ about the x-ray beam (x-axis, Fig. \ref{odt_scheme}b) such that in lab-space, the cylindrical microdomains now lay perpendicular to the field direction, along the y-axis. The field was ramped up to 6 T and the system quickly heated to 67 $^{\circ}$C, in the nematic state just below $T_{ODT}$. Time resolved measurements followed the reorientation of the cylinders from the initial horizontal to the final preferred vertical alignment, parallel to the field and z-axis. The scattering peak at q=0.7 nm $^{-1}$ from the cylindrical microdomains thus progressively shifted from concentration along the meridional line or z-axis ($\phi$=90$^{\circ}$ and 270$^{\circ}$ azimuthally) to concentration along the equatorial line or y-axis ($\phi$=180$^{\circ}$ and 360$^{\circ}$) as shown by aziumuthal intensity plots, $I(\phi,t)$, Fig. \ref{kinetics}a. This continuous microdomain rotation in a bulk nematic state stands in contrast with the non-continuous nature of director reorientation observed by time-resolved studies in smectic A thin films as recently reported \cite{brimicombe2009time} and the overall complexity of reorientation in the smectic state \cite{bras2004field}. The intensity variations of the microdomain scattering, integrated azimuthally $\pm$ 15$^{\circ}$ around $\phi=90$ and $\phi=180 ^{\circ}$ are shown in Fig. \ref{kinetics}b, with 2D SAXS data at pertinent timepoints, Fig. \ref{kinetics}c. Comparison of the initial and final timepoints indicates that the combined peak intensity is conserved. Likewise, the total scattered intensity integrated from $\phi=0$ to $2\pi$ is invariant during the experiment. This supports our earlier assertion that selective melting is not prominent here as such melting would likely have produced a transient overall intensity reduction during realignment. We therefore conclude that alignment proceeds by grain rotation as shown schematically in Fig. \ref{kinetics}c.

Exponential fits of the time traces reveal characteristic times of $\approx$ 300 and 200 min. for the $\phi=180^{\circ}$ rise and $\phi=90^{\circ}$ decrease of scattering. Consistent with this, at short times there is an initially faster loss of intensity from the 90$^{\circ}$ peak than the corresponding gain of intensity at the 180$^{\circ}$ direction. Correspondingly, at long times, the 180$^{\circ}$ peak continues to gain in intensity while loss at 90$^{\circ}$ has plateaued. These data suggest there is an intermediate state via which the system transitions from horizontal to vertically aligned cylinders. This is confirmed by the 2D data in which 4-fold symmetric chevron-like patterns are clearly observed for the BCP scattering, and to a lesser extent but still visible, for the LC smectic peaks, Fig. \ref{kinetics}c, inset (ii). To collect this data, the sample was cooled to 50 $^{\circ}$C after 250 minutes at 67 $^{\circ}$C. This provided greater segregation between the blocks and an improvement in the ability to resolve individual features which are smeared out at higher temperature. Such a display of 4-fold symmetric scattering is not unexpected. It is consistent with what is observed experimentally \cite{cohen2000deformation} and in simulation \cite{makke2012nanoscale} during the reorientation of lamellar microdomanis under tensile deformation and is clearly the result of the n=2 degeneracy with which the system can rotate to re-align with the field - i.e. microdomains can rotate to the right or the the left, with equal probability. This appears to be the first observation, however, of the expected 4-fold symmetric chevron-like state from \textit{in situ} experiments in field alignment of BCPs. Prior work by B\"{o}ker et al. \cite{boker2003electric,boker2002microscopic} observed an apparent isotropic intermediate state during electric field induced re-alignment, but it is entirely possible that this was the result of peak broadening that blurred the identification of these features in the weak segregation regime in which the experiment was conducted. The fact that the system here needed to be cooled to 50 $^{\circ}$C in order to unambiguously observe the 4-fold intermediate state underscores this point.

The timescale for reorientation is understandably large as rotational motion is severely hindered by the polymer viscosity. In general, alignment of soft mesophases by external fields is advanced on passage across the ODT as this is presumed to maximize the coupling of the field to the system \cite{darling2007dsa}. This is certainly observed empirically, but the underlying physics is a subtle combination of thermodynamic and kinetic contributions. System mobility is high near $T_{ODT}$, but the thermodynamic driving force for alignment is linked to the relevant order parameter, $\Delta\chi$ in the present case, which increases from zero only on passage through the ODT. We can construct a simple model describing the alignment kinetics as a function of temperature. We represent $\Delta\chi(T)$ for the first order N-I transition in the conventional Haller approach \cite{haller1975thermodynamic} as $\Delta\chi(T)=\Delta\chi_0\left(1-T/T_{NI}\right)^{m}$ where $\Delta\chi_0$ is the limiting value of the order parameter and $m$ is a fitting parameter capturing the steepness of the divergence and assumes a value of 0.22 in the Maier-Saupe mean-field approximation \cite{tough1983determination}. The characteristic time for alignment, $\tau$, is given by the balance of viscosity $\eta(T)$ against the field interaction energy $F(\Delta\chi(T))$. Using a simple Arrhenius model, $\eta=\eta_0\exp(E/k_B T)$ where $E$ represents the energetic barrier to flow, and $F\sim\Delta\chi$ as discussed previously. Thus we expect $\tau$ as shown in Eq.\ref{eq:alignment_timescale} and that the orientational order parameter which captures the degree of alignment evolves exponentially according to this timescale as $\langle P_2 \rangle\sim 1-\exp(-t/\tau)$ \cite{boker2003electric,schmidt2007scaling}. Any contributions from the influence of the field on the order parameter $\Delta\chi$ are small as discussed above and can be reasonably neglected.

\begin{equation}
\tau\sim\left(\frac{2\mu_0\eta_0}{\Delta\chi_0B^2}\right)\frac{\exp(E/k_B T)}{(1-T/T_{NI})^{m}}
\label{eq:alignment_timescale}
\end{equation}

\begin{figure}
\centering
\includegraphics[width=85mm, scale=1]{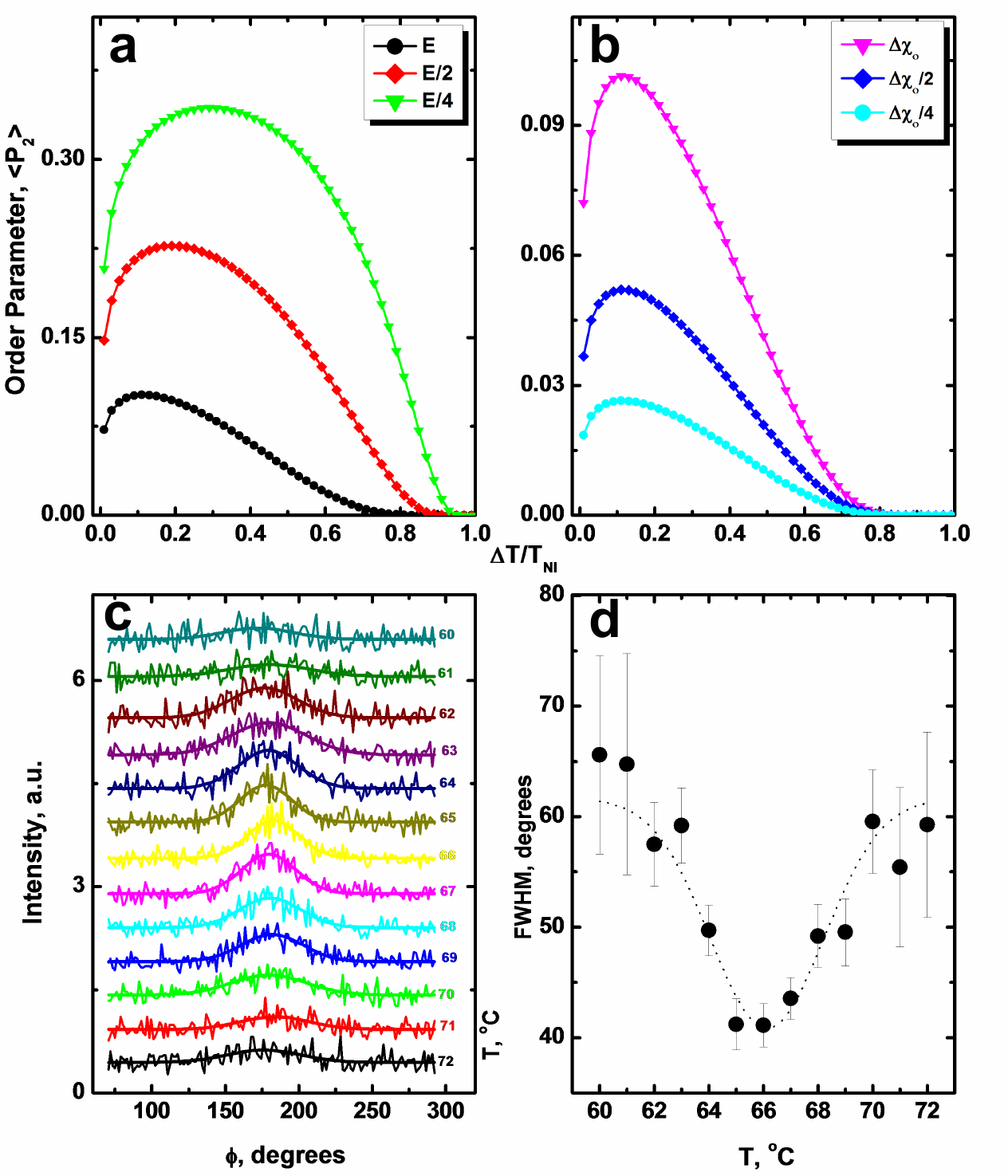}
\caption{a,b: $\langle P_2 \rangle$ dependence on annealing temperature at time $t=0.1\tau_{\tilde{T}=0.05}$; $m=0.22$, $(2\mu_0\eta_0/\Delta\chi_0B^2)=1$, $\Delta\chi_0=1$, $E/k_B T_{NI}=1.5$. c: Azimuthal intensity variation for samples annealed at different temperatures. d: FWHM of azimuthal intensity. Line is a guide to the eye.}
\label{annealing}
\end{figure}

 It is clear that for a given annealing time, there will be an optimum temperature range over which $\langle P_2\rangle$ is maximized, as shown in Fig. \ref{annealing}a,b in terms of normalized temperature $\tilde{T}=1-(T/T_{NI})$. This qualitative behavior is remarkably well captured experimentally. Samples were quenched from the disordered state to various temperatures below $T_{ODT}$ and isothermally annealed at 6 T for 1 hour. The azimuthal variation of scattered intensity from the BCP microdomains is shown in Fig. \ref{annealing}c with the temperature dependence of the FWHM of Gaussian fits (Fig. \ref{annealing}d) where smaller FWHM are correlated with higher $\langle P_2 \rangle$. FWHM is used here as an internally consistent measure of the degree of alignment in preference to estimations of $\langle P_2\rangle$ from the scattering data due to the sensitivity of this calculation to background subtraction.

\begin{figure}
\centering
\includegraphics[width=85mm, height=30mm]{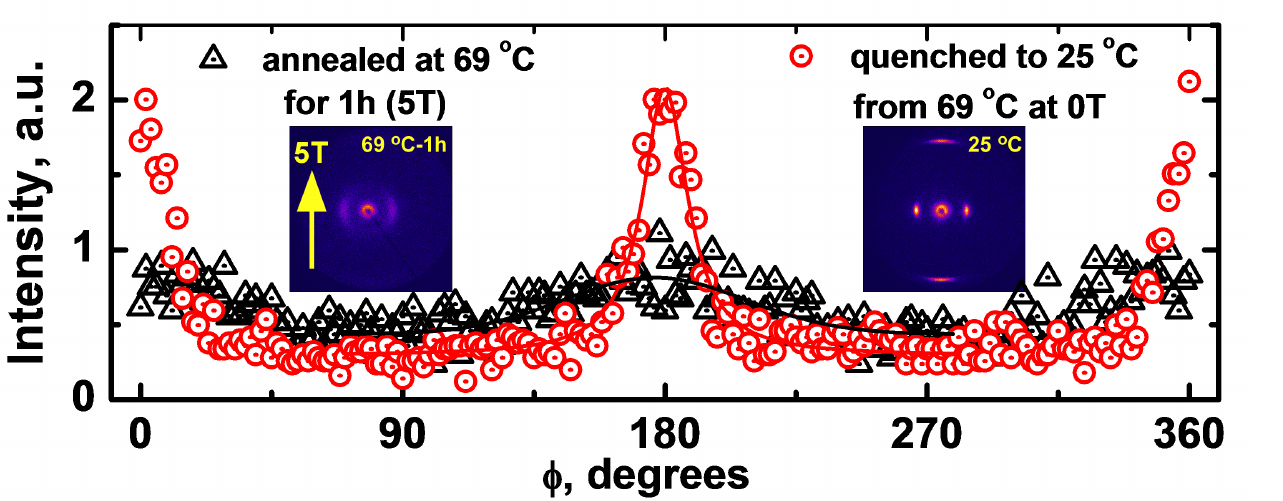}
\caption{Azimuthal intensity at 69 $^{\circ}$C and on subsequent zero-field cooling to 25 $^{\circ}$C with associated 2D SAXS data.}
\label{anneal_quench}
\end{figure}

From Fig. \ref{annealing}d it is apparent that viscosity strongly dominates in this system. The BCP has a very limited ability to respond to the field due to severe kinetic limitations starting only 6-8 $^{\circ}$C below $T_{ODT}$. This is not surprising given the location of the SmA transition to its associated large increase in viscosity over the nematic phase. The data suggest that the strong alignment observed in samples that were slowly cooled to 25 $^{\circ}$C originates only due to the residence of the system within this small temperature range and not due to the continuous action of the field down to the final temperature. This is confirmed by samples which were zero-field quenched at $\approx$ 10 $^{\circ}$C/min after an isothermal anneal for 1 hour at 5 T at 69 $^{\circ}$C. As shown in Fig. \ref{anneal_quench}, the system is only weakly ordered and aligned at 69 $^{\circ}$C, but cooling to room temperature in the absence of the field produces a sharply aligned system with similar intensities and azimuthal FWHM as obtained during continuous cooling ramps of 0.1 $^{\circ}$C/min (Fig. \ref{odt_scheme}b). The weak alignment produced during the isothermal anneal is thus sufficient to bias or template the alignment of the system as it undergoes additional ordering on cooling, even in the absence of the field. We speculate that the intervening $N-SmA$ transition may underpin this orientation refinement, in part due to a modest change in $\Delta\chi$ at the transition, but more significantly due to a coupling of the microdomain alignment with the strong smectic layer alignment as opposed to the weaker orientational order of the mesogens themselves. Further studies are clearly warranted to properly address this interesting feature.

In summary, we have provided detailed experimental \textit{in situ} data on the real-time response of LC BCPs under high magnetic fields. Our work shows that in these systems, there is no appreciable field effect on $T_{ODT}$ up to field strengths of 6 T. We can reasonably account for this given the large enthalpy associated with the ODT relative to the field interaction energy. Time resolved measurements show that slow grain rotation is active as the mechanism for alignment during isothermal annealing and that selective melting cannot play a substantive role. There is a severe kinetic limitation that arises a few $^{\circ}$C below the ODT which results in a restricted range of temperatures over which field alignment of the microstructure can be reasonably advanced. Consistent with this, we observe that the action of the field in a small window near ODT is sufficient to template very strong alignment of the system on cooling in the absence of the field and is responsible for the alignment response of the system in general. Overall, these results demonstrate new capabilities for \textit{in situ} study of BCP and soft matter physics under large magnetic fields and have important implications for the design of schemes for directed self-assembly of such systems.

\begin{acknowledgments}
The authors thank Profs. E. Thomas, F. Bates and R. Segalman for fruitful discussions, Mike Degen (Rigaku Inc.) and AMI Inc. for technical support, and gratefully acknowledge funding by NSF DMR-0847534.
\end{acknowledgments}

\bibliographystyle{apsrev}
\bibliography{insitu_alignment}

\end{document}